\documentclass[reprint,superscriptaddress,amsmath,amssymb,aps,pra]{revtex4-2}
\usepackage[colorlinks, linkcolor = ForestGreen, anchorcolor = blue, citecolor = magenta]{hyperref}
\usepackage{amsmath,amssymb,amsthm,mathtools,mathrsfs}
\usepackage[table,dvipsnames]{xcolor}
\usepackage{tikz}
\usepackage{adjustbox}
\usepackage{dsfont}
\usepackage[normalem]{ulem}
\usepackage{enumerate}
\usepackage{physics}

\theoremstyle{plain}
\newtheorem{theorem}{Theorem}
\newtheorem{lemma}{Lemma}
\newtheorem{proposition}{Proposition}

\theoremstyle{definition}
\newtheorem{definition}{Definition}
\newtheorem{remark}{Remark}
\newtheorem{example}{Example}

\newcommand{\bd}{\begin{definition}}
\newcommand{\ed}{\end{definition}}
\newcommand{\bt}{\begin{theorem}}
\newcommand{\et}{\end{theorem}}
\newcommand{\bn}{\begin{proposition}}
\newcommand{\en}{\end{proposition}}
\newcommand{\be}{\begin{equation}}
\newcommand{\ee}{\end{equation}}
\newcommand{\blem}{\begin{lemma}}
\newcommand{\elem}{\end{lemma}}
\newcommand{\bx}{\begin{example}}
\newcommand{\ex}{\end{example}}
\newcommand{\bprf}{\begin{proof}}
\newcommand{\eprf}{\end{proof}}

\DeclareMathAlphabet{\mathpzc}{OT1}{pzc}{m}{it} 
 \DeclareFontFamily{OT1}{pzc}{}
 \DeclareFontShape{OT1}{pzc}{m}{it}{ <-> s*[1.2] pzcmi7t }{}
 \DeclareMathAlphabet{\mathpzc}{OT1}{pzc}{m}{it}
 
\def\C{{{\mathbb C}}}
\newcommand{\id}{\operatorname{id}}

\def\E{\mathcal{E}}
\def\H{\mathcal{H}}

\def\J{\mathscr{J}}
\def\C{\mathbb{C}}

\begin{document}									
\preprint{APS/123-QED}

\title{Elimination of quantum causal confusion in the presence of dual states}

\author{James Fullwood}
\email{fullwood@hainanu.edu.cn}
\affiliation{School of Mathematics and Statistics, Hainan University, Haikou, Hainan Province, 570228, China}
\affiliation{Hainan International Exchange Center for Theoretical Physics, Haikou, Hainan Province, 570228, China}
\author{Weixiang Ye}
\email{wxy@hainanu.edu.cn}
\affiliation{Hainan International Exchange Center for Theoretical Physics, Haikou, Hainan Province, 570228, China}
\affiliation{Center for Theoretical Physics, School of Physics and Optoelectronic Engineering, Hainan University, Haikou, Hainan Province, 570228, China}

\date{\today}

\begin{abstract}
There is a rich class of bipartite quantum correlations between qubits which are consistent both with measurements performed in parallel and measurements performed in sequence. This phenomenon has recently been termed \emph{quantum causal confusion}, which is made mathematically precise by the notion of a dual state---a bipartite density operator whose associated correlations are causally ambiguous. A natural question is then whether or not the spatial and temporal scenarios associated with a dual state are fundamentally indistinguishable from a causal perspective. Here we show that as long as the reduced density operator of a dual state is oblique---meaning that all coefficients in its Pauli expansion are non-zero---then the underlying joint probabilities giving rise to such causally ambiguous correlations will differ in the spatial and temporal scenarios. Specifically, the temporal joint probabilities will inherently violate the no-signalling condition, thus eliminating quantum causal confusion. Moreover, as oblique states form an open dense subset of the Bloch ball, the elimination of quantum causal confusion is the norm rather than an exception. 
\end{abstract}

	\maketitle

The most fundamental task in causal inference is that of distinguishing between direct and common cause. In the context of quantum correlations, the simplest manifestation of these distinct causal scenarios takes the form of sequential measurements of a single quantum system (direct cause) and measurements performed in parallel on a joint quantum state (common cause). Moreover, as a measurement of a quantum system inherently alters its state, temporal correlations arising from sequential measurements typically differ from spatial correlations due to measurement back-action, thus providing a distinct quantum advantage for identifying quantum correlations which imply causation~\cite{FJV15,Ried_2015,Costa_2016,Costa_2018,ZPTGVF18,K_bler_2018,Meng_2024,Liu_2025,Liu_2026}. 

However, it has recently come to light that there is a rich class of bipartite quantum correlations that is consistent with both measurements performed in parallel and measurements performed in sequence~\cite{song23}. In particular, if Alice and Bob each perform Pauli measurements on a system of qubits in their respective laboratories, then the correlation data corresponding to the expectation value of the product of their measurements does not generally enable one to distinguish between direct and common cause---a situation which was termed \emph{quantum causal confusion} in Ref.~\cite{Tolotti_2026}. A natural question is then whether or not the spatial and temporal scenarios giving rise to such correlations are fundamentally indistinguishable from a causal perspective.

Quantum causal confusion is made mathematically precise by the notion of a \emph{dual state}~\cite{FuLi_2026}, which is a bipartite density operator whose correlations also admit a causal explanation~\cite{Song_2025}. In this work, we show that while the correlations encoded by a dual state are causally ambiguous, the joint probabilities which give rise to such correlations will in general differ in the common cause and direct cause scenarios. We emphasize that our treatment of such direct cause scenarios is quite general, as we allow the intermediate dynamics between sequential measurements to include arbitrary system-environment interactions modeled by a quantum channel. Moreover, in such a case we show that the crucial distinction between direct and common cause lies in the no-signalling condition for the associated joint probabilities, which ensures that Alice's measurement outcomes are independent of Bob's and vice-versa. In particular, while in a spatial scenario (common cause) it follows from the Born rule that the underlying joint probabilities always satisfy no-signalling, in a temporal scenario (direct cause) the associated joint probabilities may or may not satisfy no-signalling. As such, a violation of no-signalling implies direct cause, thus eliminating quantum causal confusion in the presence of dual states. 

In our main result, we show that if a reduced density operator of a dual state is \emph{oblique}---meaning its coefficients in a Pauli expansion are all non-zero---then the joint probabilities which give rise to the correlations encoded by the dual state will necessarily differ in common cause and direct cause scenarios. Specifically, we show that in the case of direct cause there will necessarily exist a pair of Pauli measurements whose joint probabilities violate the no-signalling condition, thus eliminating causal confusion. Furthermore, we show that the oblique condition is generically satisfied for dual states, implying that the elimination of quantum causal confusion is the norm rather than an exception. Therefore, quantum causal confusion in the presence of dual states may generally be attributed to information loss due to a post-processing of Alice and Bob's individual measurement outcomes to form joint expectation values, inherently masking whether or not the no-signalling condition is violated.

\section{The physical setup}

Suppose Alice and Bob perform an experiment, where each run of the experiment consists of Alice and Bob each performing a local measurement on a single qubit~\footnote{While everything here immediately generalizes to systems of $n$-qubits, we focus on the case of a single qubit for clarity of exposition.}. We denote Alice's qubit by $A$ and Bob's qubit by $B$. We also assume that for each run of the experiment, Alice and Bob measure an observable which is chosen at random from the Pauli basis $\{\mathds{1},\sigma_x,\sigma_y,\sigma_z\}$ of qubit observables (so that $\mathds{1}$ corresponds to the `empty measurement'). 

If for each run of the experiment Alice and Bob perform a measurement on a shared joint state $\rho_{AB}$, then we may assume without loss of generality that Alice and Bob's measurements are spacelike separated for each run of the experiment. This will be referred to as the \emph{spatial scenario} associated with the joint state $\rho_{AB}$. In such a case, the correlations between Alice and Bob's measurements may be explained in terms of the common cause corresponding to a preparation of the joint state $\rho_{AB}$. 

If, on the other hand, each run of the experiment consists of Alice performing a measurement on a fixed state $\rho_A$, and then sending the updated state via a quantum channel $\E$ to Bob who subsequently performs a measurement on the output of the channel, then Alice and Bob's qubits are necessarily timelike separated for each run of the experiment. This will be referred to as the \emph{temporal scenario} associated with the state-channel pair $(\rho_A,\E)$. In such a case, Alice's measurement outcome will in general have a direct causal effect on Bob's measurement outcome due to the measurement back-action resulting from state disturbance.

\begin{figure*}[htpb]
    \centering
    \includegraphics[width=0.98\textwidth]{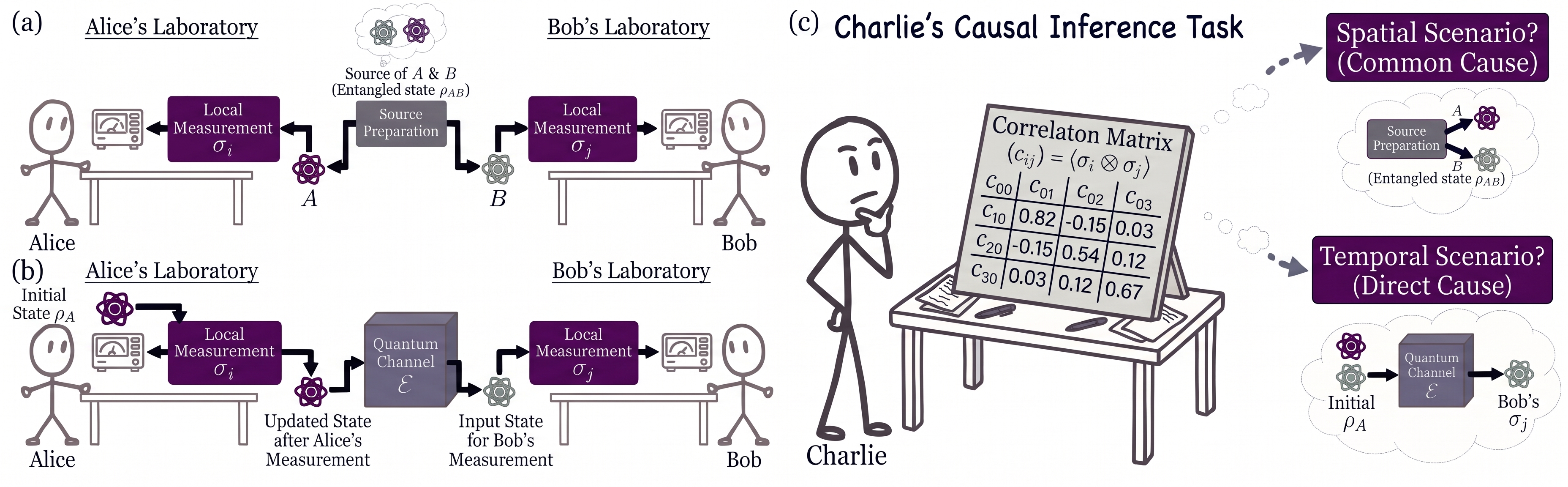}
    \caption{
    \label{F1}
    \textbf{Quantum causal inference.} (a) In the spatial scenario (common cause), Alice and Bob perform Pauli measurements $\sigma_i$ and $\sigma_j$ on a shared joint state $\rho_{AB}$. (b) In the temporal scenario (direct cause), Alice prepares an initial state $\rho_A$ and performs a Pauli measurement $\sigma_i$. She then sends the updated state via a quantum channel $\E$ to Bob, who subsequently performs a Pauli measurement $\sigma_j$. (c) Charlie, who is given the correlation data $(c_{ij})=\langle \sigma_i\otimes \sigma_j \rangle$ of Alice and Bob's measurements, is assigned the task of inferring whether the data was generated from a spatial or temporal scenario.}
\end{figure*}

Now suppose that after the experiment, numerical data is gathered in the form of a $4\times 4$ correlation matrix $(c_{ij})=\langle \sigma_i\otimes \sigma_j \rangle$, where $\langle \sigma_i\otimes \sigma_j \rangle$ is the average of the product of Alice and Bob's measurements whenever Alice measured $\sigma_i$ and Bob measured $\sigma_j$. Furthermore, suppose that the data $c_{ij}$ is given to Charlie, who is assigned the task of inferring whether the data was generated by a spatial or temporal scenario (as depicted in Figure~\ref{F1}). To this end, Charlie takes the data $c_{ij}$ and forms the associated \emph{correlation operator} $\tau_{AB}$ given by
\[
\tau_{AB}=\frac{1}{4}\sum_{i,j}c_{ij}\sigma_i\otimes \sigma_j\, .
\]

If the data $c_{ij}$ was generated by a spacelike scenario, then it necessarily follows that $\tau_{AB}$ is a density operator, and thus is positive semi-definite. Therefore, if the correlation operator $\tau_{AB}$ has a negative eigenvalue, then Charlie can be sure that the data $c_{ij}$ was generated by a temporal scenario~\cite{FJV15}. However, if $\tau_{AB}$ is in fact positive semi-definite, then Charlie cannot infer the causal origin of the correlation data $c_{ij}$ from the spectrum of $\tau_{AB}$. In particular, it has been shown in Refs.~\cite{ZPTGVF18,song23} that there exists correlation data $c_{ij}$ that is consistent with both spatial and temporal scenarios, thus resulting in \emph{quantum causal confusion} for Charlie. 

\section{The underlying theory}

To further investigate the origins of quantum causal confusion, we first recall the basic theory underlying the spatial and temporal scenarios outlined in the previous section. For this, let $\H_A\cong \C^{2}$ and $\H_B\cong \C^{2}$ denote the Hilbert spaces associated with Alice and Bob's qubits, and let $AB$ denote the composite system. Given $X\in \{A,B,AB\}$, the algebra of linear operators on $\H_X$ will be denoted by $\mathcal{L}(\H_X)$, and the space of density operators on $\H_X$ will be denoted by $\mathfrak{S}(X)\subset \mathcal{L}(\H_X)$. A linear map $\E:\mathcal{L}(\H_A)\to \mathcal{L}(\H_B)$ which is completely positive and trace-preserving (CPTP) will be referred to as a \emph{quantum channel}. We let $\{\sigma_0,\sigma_1,\sigma_2,\sigma_3\}$ denote the standard Pauli basis for Hermitian operators on $\C^2$ (where $\sigma_0=\mathds{1}$, $\sigma_1=\sigma_x$, $\sigma_2=\sigma_y$ and $\sigma_3=\sigma_z$), and we let $\Pi_{i}^{\pm}$ denote the spectral projectors associated with $\sigma_{i}\in \{\sigma_1,\sigma_2,\sigma_3\}$, so that $\sigma_{i}=\Pi_{i}^{+}-\Pi_{i}^{-}$ and $\Pi_{i}^{+}+\Pi_{i}^{-}=\mathds{1}$ for all $i\in \{1,2,3\}$. 

\subsection{The spatial scenario}

In the spatial scenario outlined in the previous section, Alice and Bob perform local measurements of Pauli observables on a shared joint state $\rho_{AB}$. In such a case, it follows from the Born rule that the probability $\mathbb{P}(x,y|i,j)$ that Alice gets outcome $x\in \{+,-\}$ and Bob gets outcome $y\in \{+,-\}$ provided that Alice measures $\sigma_i$ and Bob measures $\sigma_j$ is given by
\be \label{JNTXPRBSP37}
\mathbb{P}(x,y|i,j)=\Tr(\rho_{AB}\, \Pi_i^x\otimes \Pi_j^y)\quad \forall i,j\in \{1,2,3\} \, . 
\ee
We note that the cases $i=0$ and $j=0$ correspond to Alice and Bob performing the `empty measurement' respectively, with which we associate the constant outcome $+$. For example, when Alice performs the empty measurement ($i=0$) we have $\mathbb{P}(-,y|0,j)=0$ and $\mathbb{P}(+,y|0,j)=\Tr(\rho_B\Pi_j^y)$ for all $y\in \{+,-\}$ and for all $j\in \{1,2,3\}$, where $\rho_B=\Tr_A(\rho_{AB})$. It then follows that if $\langle \sigma_{i}\otimes \sigma_{j} \rangle$ denotes the expectation value of the product of Alice and Bob's measurements when Alice measures $\sigma_{i}$ and Bob measures $\sigma_{j}$, then
\begin{align*}
\langle \sigma_{i}\otimes \sigma_{j} \rangle&=\sum_{x,y\in \{+,-\}} (-1)^{1+\delta_{xy}}\mathbb{P}(x,y|i,j) \\
&=\Tr(\rho_{AB}\, \sigma_{i}\otimes \sigma_{j})\, ,
\end{align*}
thus by properties of Pauli matrices we have
\be \label{PAULIEXP57}
\rho_{AB}=\frac{1}{4}\sum_{i,j=0}^{3}\langle \sigma_{i}\otimes \sigma_{j} \rangle \sigma_i\otimes \sigma_j\, .
\ee

While the correlations between Alice and Bob's measurements may exhibit non-local characteristics, Alice's individual measurement outcomes are completely independent of which measurement Bob performs and vice-versa. Such a property is referred to as the \emph{no-signalling} condition, which is captured mathematically by the requirement that the following marginal conditions hold:
\begin{enumerate}[i.]
\item \label{NSBTA}
$\sum_{y\in \{+,-\}} \mathbb{P}(x,y|i,j)=\sum_{y\in \{+,-\}} \mathbb{P}(x,y|i,j')$ for all $i,j,j'\in \{0,1,2,3\}$ and for all $x\in \{+,-\}$.
\item \label{NSATB}
$\sum_{x\in \{+,-\}} \mathbb{P}(x,y|i,j)=\sum_{x\in \{+,-\}} \mathbb{P}(x,y|i',j)$ for all $i,i',j\in \{0,1,2,3\}$ and for all $y\in \{+,-\}$.
\end{enumerate}
In particular, condition \ref{NSBTA} may be referred to as ``no-signalling from Bob to Alice", as it ensures that Alice's individual measurement outcomes are independent of Bob's. Similarly, condition \ref{NSATB} may be referred to as ``no-signalling from Alice to Bob", as it ensures that Bob's individual measurement outcomes are independent of Alice's. The fact that both conditions \ref{NSBTA} and \ref{NSATB} are satisfied in such a spatial scenario immediately follows from the Born rule~\eqref{JNTXPRBSP37}. 

\subsection{The temporal scenario}

In the temporal scenario, Alice performs a measurement on a fixed state $\rho_A$, and then sends the updated state via a quantum channel $\E:\mathcal{L}(\H_A)\to \mathcal{L}(\H_B)$ to Bob who subsequently measures the output of the channel. Such a temporal scenario will be labeled by the data consisting of the state-channel pair $(\rho_A,\E)$. According to the L\"{u}ders update rule~\cite{Lu06}, the probability $\mathbb{P}(x,y|i,j)$ that Alice gets outcome $x\in \{+,-\}$ and Bob gets outcome $y\in \{+,-\}$ provided that Alice measures $\sigma_i$ and Bob measures $\sigma_j$ is given by
\be \label{LUDEXSP67}
\mathbb{P}(x,y|i,j)=\Tr(\E(\Pi_i^x\rho_A \Pi_i^x)\Pi_j^y)\quad \forall i,j\in \{1,2,3\}\, .
\ee
As in the spatial scenario, the cases $i=0$ and $j=0$ correspond to the `empty measurement', which we associate with the constant outcome $+$. For example, when Alice performs the empty measurement ($i=0$), the update rule yields $\mathbb{P}(-,y|0,j)=0$ and $\mathbb{P}(+,y|0,j)=\Tr(\E(\rho_A)\Pi_j^y)$ for all $y\in \{+,-\}$ and $j\in \{1,2,3\}$. Conversely, if Bob performs the empty measurement ($j=0$), the trace-preserving property of $\E$ ensures that $\mathbb{P}(x,-|i,0)=0$ and $\mathbb{P}(x,+|i,0)=\Tr(\rho_A\Pi_i^x)$ for all $x\in \{+,-\}$ and $i\in \{1,2,3\}$.

A distinctive feature of the temporal probability rule \eqref{LUDEXSP67} is that the projector $\Pi_i^x$ labels a measurement outcome while also taking an active role in updating the state upon Alice's measurement. This may result in measurement back-action, which in general will have direct causal influence on Bob's measurement outcomes. In particular, the no-signalling condition from Alice to Bob (condition \ref{NSATB}) will not necessarily hold for a given state-channel pair $(\rho_A,\E)$. We note that setting $i'=0$ in \ref{NSATB} results in the condition
\be \label{eq:NSIT_def}
\sum_{x \in \{+,-\}} \Tr( \E(\Pi_i^x \rho_A \Pi_i^x) \Pi_j^y ) = \Tr( \E(\rho_A) \Pi_j^y )
\ee
for all $i,j\in \{0,1,2,3\}$ and for all outcomes $y \in \{+,-\}$, which is often referred to as the \emph{no-signalling in time} condition~\cite{KB_2013}. It is immediate that if $[\rho_A, \Pi_i^x]=0$ for all $i\in \{0,1,2,3\}$ and for all $x\in \{+,-\}$, or if there exists a state $\sigma\in \mathfrak{S}(B)$ such that $\E(\omega)=\sigma$ for all $\omega\in \mathfrak{S}(A)$, then no-signalling in time is necessarily satisfied. The no-signalling in time condition is then a weakening of the full no-signalling conditions \ref{NSBTA} and \ref{NSATB}, as it only requires that Bob's probabilities are independent of whether or not Alice performed a given measurement. As such, a violation of no-signalling in time implies a violation of no-signalling. Moreover, violations of no-signalling in time imply a violation of \emph{passive realism}~\cite{FYW_2026}, which is a scale-independent formulation of macroscopic realism founded upon the assumptions that physical systems have well-defined properties independent of whether or not such properties are measured, and that it is possible to passively measure such properties of a system without altering the subsequent dynamics of the system.

On the other hand, we note that no-signalling from Bob to Alice (condition \ref{NSBTA}) is always satisfied for the temporal joint probabilities given by \eqref{LUDEXSP67}. In particular, there are no retrocausal effects in such a temporal scenario, as Alice's measurement outcomes are completely independent of Bob's. This is often referred to as the ``arrow of time" (or ``causality") condition in the literature~\cite{Leggett_2002,Chiribella_2011,Kofler_2013,Clemente_2015,Vitag_2023}.

When Alice first measures $\sigma_i$ and then Bob measures $\sigma_j$, the expectation value $\langle \sigma_i\otimes \sigma_j \rangle$ of the product of Alice and Bob's measurements is then given by 
\[
\langle \sigma_{i}\otimes \sigma_{j} \rangle=\sum_{x,y \in \{+,-\}} (-1)^{1+\delta_{xy}}\mathbb{P}(x,y|i,j) \, .
\]  
Contrary to the spatial case, there does not necessarily exist a bipartite density operator $\rho_{AB}$ such that $\Tr(\rho_{AB}\, \sigma_i\otimes \sigma_j)$ coincides with the expectation values $\langle \sigma_{i}\otimes \sigma_{j} \rangle$ for all $i$ and $j$. However, in the spirit of Eq.~\eqref{PAULIEXP57} we can take the temporal expectation values $\langle \sigma_{i}\otimes \sigma_{j} \rangle$ and form the operator $\varrho_{AB}$ given by 
\[
\varrho_{AB}=\frac{1}{4}\sum_{i,j=0}^{3}\langle \sigma_{i}\otimes \sigma_{j} \rangle\sigma_i\otimes \sigma_j\, ,
\]
which is the \emph{pseudo-density operator} associated with the state-channel pair $(\rho_A,\E)$~\cite{FJV15}. The prefix `pseudo' indicates that the operator $\varrho_{AB}$ will not in general be positive semi-definite. Nevertheless, by properties of Pauli matrices it follows that for all $i$ and $j$ we have
\[
\langle \sigma_{i}\otimes \sigma_{j} \rangle=\Tr(\varrho_{AB}\, \sigma_i\otimes \sigma_j)\, ,
\]
so that the pseudo-density operator $\varrho_{AB}$ encodes temporal expectation values in precisely the same manner that a density operator encodes spatial expectation values. 

Remarkably, the pseudo-density operator $\varrho_{AB}$ associated with a state-channel pair $(\rho_A,\E)$ admits the basis-independent formula
\be \label{ANTXCMT87}
\varrho_{AB}=\frac{1}{2}\big\{\rho_A\otimes \mathds{1}_B\, ,\J[\E]\big\}\, ,
\ee
where the bracket $\{\cdot\, ,\cdot\}$ denotes the anticommutator and $\J[\E]=\sum_{i,j}\dyad{i}{j}\otimes \E(\dyad{j}{i})$ is the \emph{Jamio\l kowski operator} associated with the quantum channel $\E$. We note that the Jamio\l kowski operator $\J[\E]$ is a partial transpose of the \emph{Choi matrix} $\mathscr{C}[\E]=\sum_{i,j}\dyad{i}{j}\otimes \E(\dyad{i}{j})$~\cite{Ch75}. For ease of notation, we denote the right-hand side of Eq.~\eqref{ANTXCMT87} by $\E\star \rho_A$ moving forward. 

\subsection{Dual states}

A bipartite density operator $\rho_{AB}\in \mathfrak{S}(AB)$ is said to be a \emph{dual state} if and only if there exists a quantum channel $\E:\mathcal{L}(\H_A)\to \mathcal{L}(\H_B)$ such that 
\be \label{DUALSXT71}
\rho_{AB}=\E\star \rho_A\, , 
\ee
where $\rho_A=\Tr_B(\rho_{AB})$, and we recall that the $\star$-product is given by $\E\star \rho_A=\frac{1}{2}\{\rho_A\otimes \mathds{1}_B\, ,\J[\E]\}$. In other words, $\rho_{AB}$ is a dual state if and only if it is also a pseudo-density operator. Dual states then admit a dual interpretation as encoding the correlations in a spatial scenario associated with the density operator $\rho_{AB}$, while also encoding the correlations in a temporal scenario associated with the state-channel pair $(\rho_A,\E)$. It then follows that the correlations $\langle \sigma_i\otimes \sigma_j\rangle$ encoded by a dual state are causally ambiguous~\cite{Lorenz_2026}, resulting in \emph{quantum causal confusion}~\cite{Tolotti_2026}.

Mathematically, one can approach the phenomenon of quantum causal confusion from two distinct routes. First, given a bipartite density operator $\rho_{AB}$, one may ask whether or not there exists necessary and sufficient conditions for the existence of a quantum channel $\E$ such that Eq.~\eqref{DUALSXT71} holds. Second, given a state-channel pair $(\rho_A,\E)$, one may ask whether or not there exist necessary and sufficient conditions for the positive semi-definiteness of $\E\star \rho_A$, which implies the existence of a density operator $\rho_{AB}$ such that Eq.~\eqref{DUALSXT71} holds. Such necessary and sufficient conditions were established for the former case in Ref.~\cite{Song_2025}, where it was shown that a density operator $\rho_{AB}$ is a dual state if and only if $(\mathcal{D}\otimes \id_B)(\rho_{AB})$ has positive partial transpose (PPT), where $\mathcal{D}$ is a dephasing map associated with an eigenbasis of $\rho_A$. For the latter case, it was shown in Ref.~\cite{Tolotti_2026} that $\E\star \rho_A$ is positive semi-definite if and only if $\E$ is a PPT channel, which means that the Choi matrix $\mathscr{C}[\E]$ has positive partial transpose. In the context of single qubit channels, the PPT condition is equivalent to the channel $\E$ being \emph{entanglement breaking}~\cite{Horodecki_2003}. Furthermore, it was established in Refs.~\cite{FuLi_2026,Song_2025} that all PPT states are in fact dual states. In particular, all separable states are dual states.

\section{The main result}

Let $\rho_{AB}=\E\star \rho_A$ be a dual state, so that the correlations $\langle \sigma_i\otimes \sigma_j\rangle=\Tr(\rho_{AB}\, \sigma_i\otimes \sigma_j)$ may be viewed either as arising from a spatial scenario associated with the state $\rho_{AB}$, or from a temporal scenario associated with the state-channel pair $(\rho_A,\E)$. It then follows that one is not able to distinguish the spatial and temporal scenarios associated with a dual state from the correlation data alone, thus resulting in quantum causal confusion. A natural question is then whether or not the spatial and temporal scenarios associated with a dual state are fundamentally indistinguishable from a casual perspective. Our main result states that in general, the joint probabilities associated with the spatial and temporal scenarios will necessarily differ, thus eliminating quantum causal confusion. 

For the statement of our main result, we denote the joint probabilities in the spatial scenario as given by the Born rule \eqref{JNTXPRBSP37} by $\mathbb{P}_{\mathrm{space}}(x,y|i,j)$, while the joint probabilities in the temporal scenario as given by the L\"{u}ders rule \eqref{LUDEXSP67} will be denoted by $\mathbb{P}_{\mathrm{time}}(x,y|i,j)$. Given a state-channel pair $(\rho_A,\E)$, the state $\rho_A$ will be referred to as \emph{oblique} if and only if its Pauli expansion $\rho_A=\frac{1}{2}(\mathds{1}+r_1\sigma_1+r_2\sigma_2+r_3\sigma_3)$ is such that $r_1r_2r_3\neq 0$, and $\E$ will be referred to as a \emph{discard-and-prepare channel} if and only if there exists a state $\sigma\in \mathfrak{S}(B)$ such that $\E(\omega)=\sigma$ for all $\omega\in \mathfrak{S}(A)$. Our main result is then given by the following:

\bt \label{thm:prob_divergence}
Let $\rho_{AB} = \E \star \rho_A$ be a dual state with $\rho_A$ an oblique state and $\E$ not a discard-and-prepare channel. Then there exist Pauli observables $\sigma_i$ and $\sigma_j$ with $i,j \in \{1,2,3\}$ such that 
\be \label{NEQ77}
\mathbb{P}_{\mathrm{space}}(x,y|i,j) \neq \mathbb{P}_{\mathrm{time}}(x,y|i,j)
\ee
for some outcomes $x,y \in \{+,-\}$.
\et 

The idea behind the proof is to show that under the hypotheses of Theorem~\ref{thm:prob_divergence}, there must necessarily exist $i,j \in \{1,2,3\}$ such that the no-signalling in time condition \eqref{eq:NSIT_def} is violated. Since the spatial probabilities $\mathbb{P}_{\mathrm{space}}(x,y|i,j)$ always satisfy no-signalling, such a violation then implies the non-equality \eqref{NEQ77} for some outcomes $x,y \in \{+,-\}$. Before giving the full proof, we first prove a lemma. For this, we note that the no-signalling condition \eqref{eq:NSIT_def} for a measurement of $\sigma_i$ followed by $\sigma_j$ combined with our notation $\mathbb{P}_{\mathrm{time}}(x,y|i,j)$ for the associated joint temporal probabilities reads as 
\be \label{eq:NSIT_condition}
\sum_{x \in \{+,-\}} \mathbb{P}_{\mathrm{time}}(x,y|i,j) = \Tr( \E(\rho_A) \Pi_j^y ) \qquad \forall y \in \{+,-\}\, .
\ee

\blem \label{lem:NSIT_equivalence}
Let $(\rho_A,\E)$ be a state-channel pair. Then the temporal joint probabilities $\mathbb{P}_{\mathrm{time}}(x,y|i,j)$ satisfy the no-signalling in time (NSIT) condition \eqref{eq:NSIT_condition} for all $i,j \in \{0,1,2,3\}$ if and only if $r_k \E(\sigma_k) = 0$ for all $k \in \{1,2,3\}$, where $r_k = \Tr(\rho_A \sigma_k)$ are the coefficients in the Pauli expansion of $\rho_A$.
\elem

\bprf
We first note that for the trivial measurement $i=0$ (where $\Pi_0^+ = \mathds{1}$ and $\Pi_0^- = 0$), the left-hand side of Eq.~\eqref{eq:NSIT_condition} reduces to $\Tr(\E(\rho_A)\Pi_j^y)$, which trivially equals the right-hand side. Similarly, for $j=0$, both sides evaluate to $\Tr(\E(\rho_A)) = 1$ due to the trace-preserving property of $\E$. Thus, the NSIT condition is inherently satisfied whenever $i=0$ or $j=0$. Consequently, the condition holds for all $i,j \in \{0,1,2,3\}$ if and only if it holds for the active Pauli measurements $i,j \in \{1,2,3\}$.

For $i,j \in \{1,2,3\}$, by the L\"{u}ders update rule given in Eq.~\eqref{LUDEXSP67}, Bob's temporal marginal is evaluated as
\[
\sum_{x \in \{+,-\}} \mathbb{P}_{\mathrm{time}}(x,y|i,j) = \Tr\Big( \E\Big( \sum_{x \in \{+,-\}} \Pi_i^x \rho_A \Pi_i^x \Big) \Pi_j^y \Big) \, .
\]
Expanding the initial state as $\rho_A = \frac{1}{2}(\mathds{1} + \sum_{k=1}^3 r_k \sigma_k)$, and using the fact that $\Pi_i^{\pm} = \frac{1}{2}(\mathds{1} \pm \sigma_i)$, we find that 
\[
\sum_{x \in \{+,-\}} \Pi_i^x \rho_A \Pi_i^x = \frac{1}{2}(\rho_A + \sigma_i \rho_A \sigma_i) = \rho_A - \frac{1}{2}\sum_{k \neq i} r_k \sigma_k \, .
\]
Substituting this into the NSIT condition \eqref{eq:NSIT_condition} yields
\[
\Tr\Big( \E\Big(\rho_A - \frac{1}{2}\sum_{k \neq i} r_k \sigma_k\Big) \Pi_j^y \Big) = \Tr( \E(\rho_A) \Pi_j^y ) \, ,
\]
which simplifies directly to
\[
\Tr\Big( \E\Big(\sum_{k \neq i} r_k \sigma_k\Big) \Pi_j^y \Big) = 0 \qquad \forall y \in \{+,-\} \, .
\]
Since $\Pi_j^{\pm} = \frac{1}{2}(\mathds{1} \pm \sigma_j)$ and $\E$ is a trace-preserving map, this condition reduces strictly to the orthogonality relation
\[
\Tr\Big( \E\Big(\sum_{k \neq i} r_k \sigma_k\Big) \sigma_j \Big) = 0 \qquad \forall i,j \in \{1,2,3\} \, .
\]
Since the Pauli matrices $\{\sigma_1, \sigma_2, \sigma_3\}$ form an orthogonal basis for the space of traceless Hermitian operators on $\C^2$, this inner product vanishes for all $j \in \{1,2,3\}$ if and only if $\E(\sum_{k \neq i} r_k \sigma_k) = 0$ for all $i \in \{1,2,3\}$. 

Expanding this over the three possible values of $i$ yields the homogeneous linear system
\begin{align*}
r_2 \E(\sigma_2) + r_3 \E(\sigma_3) &= 0 \\
r_1 \E(\sigma_1) + r_3 \E(\sigma_3) &= 0 \\
r_1 \E(\sigma_1) + r_2 \E(\sigma_2) &= 0 \, .
\end{align*}
Adding the first and third equations and subtracting the second yields $2r_2 \E(\sigma_2) = 0$. By symmetry, this implies $r_k \E(\sigma_k) = 0$ for all $k \in \{1,2,3\}$, which establishes the equivalence and completes the proof.
\eprf

\bprf[Proof of Theorem~\ref{thm:prob_divergence}]
Suppose for the sake of contradiction that $\mathbb{P}_{\mathrm{space}}(x,y|i,j) = \mathbb{P}_{\mathrm{time}}(x,y|i,j)$ for all sequential Pauli measurements $\sigma_i$ and $\sigma_j$ with $i,j \in \{0,1,2,3\}$ and for all outcomes $x,y \in \{+,-\}$. As established in Section II.A, the spatial distribution $\mathbb{P}_{\mathrm{space}}$ inherently satisfies the full no-signalling condition, which guarantees that Bob's marginal probability is completely independent of Alice's measurement choice. Consequently, under our assumption of equivalence, the temporal distribution $\mathbb{P}_{\mathrm{time}}$ must satisfy the no-signalling in time (NSIT) condition \eqref{eq:NSIT_condition} for all $i,j \in \{0,1,2,3\}$. 

By Lemma~\ref{lem:NSIT_equivalence}, this satisfaction of the NSIT condition implies that $r_k \E(\sigma_k) = 0$ for all $k \in \{1,2,3\}$, where $r_k$ are the coefficients in the Pauli expansion of the initial state $\rho_A$. Since $\rho_A$ is an oblique state by assumption, its Pauli coefficients are all non-zero, so that $r_x r_y r_z \neq 0$. This forces the condition $\E(\sigma_k) = 0$ for all $k \in \{1,2,3\}$, which by linearity implies that the quantum channel $\E$ annihilates all traceless observables. Since $\E$ is also trace-preserving, it necessarily follows that $\E(X) = \Tr(X)\sigma_B$ for some fixed state $\sigma_B \in \mathfrak{S}(B)$. It then follows that $\E$ is a discard-and-prepare channel, contradicting our initial assumption that $\E$ is not a discard-and-prepare channel. Therefore, we conclude that $\mathbb{P}_{\mathrm{space}}(x,y|i,j) \neq \mathbb{P}_{\mathrm{time}}(x,y|i,j)$ for some $i,j \in \{0,1,2,3\}$ and outcomes $x,y \in \{+,-\}$. Finally, since the spatial and temporal distributions trivially agree for the empty measurements where $i=0$ or $j=0$, this non-equality must necessarily occur for active Pauli measurements where $i,j \in \{1,2,3\}$.
\eprf

\begin{remark} \label{rem:generic_oblique}
We note that the oblique condition required by Theorem~\ref{thm:prob_divergence} is generically satisfied for any given dual state $\rho_{AB} = \E \star \rho_A$. In particular, as a single-qubit state $\rho_A$ fails to be oblique if and only if at least one coefficient in its Pauli expansion vanishes, the Bloch vector of a non-oblique state necessarily lies within one of the three principal coordinate planes. As the union of these planes constitutes a closed set of Lebesgue measure zero, the set of oblique states is an open dense subset of the Bloch ball. Consequently, a randomly chosen initial state $\rho_A$ will be oblique almost surely, thus the elimination of causal confusion in the presence of dual states is the norm rather than an exception.
\end{remark}

\begin{remark}
An immediate corollary of Theorem~\ref{thm:prob_divergence} is that if $\rho_{AB}=\E\star \rho_A$ is a dual state such that $\mathbb{P}_{\mathrm{space}}(x,y|i,j)=\mathbb{P}_{\mathrm{time}}(x,y|i,j)$ for all $x,y\in \{+,-\}$ and for all $i,j\in \{0,1,2,3\}$, then it necessarily follows that either $\rho_A$ is not oblique, or $\E$ is a discard-and-prepare channel. In the case that $\E$ is a discard-and-prepare channel, it follows that $\rho_{AB}=\rho_A\otimes \sigma_B$, where $\sigma_B\in \mathfrak{S}(B)$ is the unique state of $B$ such that $\E(\omega)=\sigma_B$ for all $\omega\in \mathfrak{S}(A)$. In other words, causal confusion persists only if the measure-zero condition of non-obliqueness for $\rho_A$ is satisfied, or if the intermediate dynamics in the associated temporal scenario completely annihilate any causal influence of Alice's measurements on Bob's outcomes. 
\end{remark}

\begin{remark} \label{rem:channel_reconstruction}
Suppose Charlie successfully eliminates causal confusion in the presence of a dual state $\rho_{AB}=\E\star \rho_A$ by identifying a violation of no-signalling, thus inferring direct cause. A natural question is whether Charlie can also infer the channel $\E$ governing the dynamics between measurements. By definition, the dual state $\rho_{AB}$ satisfies the relation $2\rho_{AB} = \{\rho_A \otimes \mathds{1}_B, \J[\E]\}$, which takes the form of a Sylvester equation for the unknown Jamio\l{}kowski operator $\J[\E]$. Such an equation admits a unique solution if and only if $\rho_A \otimes \mathds{1}_B$ and $-\rho_A \otimes \mathds{1}_B$ have disjoint spectra~\cite{Datta}. Since $\rho_A$ is positive semi-definite, this condition is satisfied if and only if $\rho_A$ is of full rank, which for a single qubit implies that $\rho_A$ is a mixed state. Therefore, provided $\rho_A$ is both oblique (to eliminate causal confusion) and of full rank, Charlie can not only infer the existence of a direct cause, but he can also uniquely solve the associated Sylvester equation for $\J[\E]$. Since the mapping $\E \mapsto \J[\E]$ is a linear isomorphism, Charlie's exact knowledge of the Jamio\l{}kowski operator allows him to explicitly reconstruct the underlying quantum channel $\E$.
\end{remark}

We now provide an explicit example of the elimination of quantum causal confusion.

\bx \label{ex:causal_confusion_resolution}
Consider the state-channel pair $(\rho_A, \E)$, where $\rho_A$ is the oblique state given by
\[
\rho_A = \frac{1}{2}\left(\mathds{1} + \frac{1}{3}\sigma_x + \frac{1}{3}\sigma_y + \frac{1}{3}\sigma_z\right)\, ,
\]
and $\E$ is a depolarizing channel given by 
\[
\E(X) = p X + (1-p)\Tr(X)\frac{\mathds{1}}{2} \, ,
\]
where we set $p=1/5$. Since $p \le 1/3$, the channel $\E$ is entanglement-breaking~\cite{Horodecki_2003} (and thus PPT), which ensures that $\rho_{AB} = \E \star \rho_A$ is a dual state~\cite{Tolotti_2026}. 

As an illustration of the elimination of causal confusion in the presence of the dual state $\rho_{AB}$ (as ensured by Theorem~\ref{thm:prob_divergence}), consider the case when Alice measures $\sigma_x$ ($i=1$) and Bob measures $\sigma_y$ ($j=2$). In the spatial scenario associated with $\rho_{AB}$, Bob's marginal probability of obtaining the outcome $y=+$ is computed as  
\[
\sum_{x \in \{+,-\}} \mathbb{P}_{\mathrm{space}}(x,+|1,2)=\Tr(\rho_B\Pi_2^+)=\frac{8}{15} \, .
\]
Conversely, in the temporal scenario associated with the state-channel pair $(\rho_A,\E)$, measurement back-action from Alice's $\sigma_x$ measurement non-selectively annihilates the $\sigma_y$ and $\sigma_z$ components of the initial state prior to the action of the channel $\E$. Bob's temporal marginal thus evaluates to
\begin{align*}
\sum_{x \in \{+,-\}} \mathbb{P}_{\mathrm{time}}(x,+|1,2)& = \Tr\Big(\E\Big(\sum_{x \in \{+,-\}} \Pi_1^x \rho_A \Pi_1^x\Big)\Pi_2^+\Big) \\
& = \Tr\Big(\E\Big(\frac{1}{2}\mathds{1} + \frac{1}{6}\sigma_x\Big)\Pi_2^+\Big) = \frac{1}{2} \, .
\end{align*}
Since $8/15 \neq 1/2$, the temporal joint probabilities violate the no-signalling in time condition. 

To see how this forces a disparity at the level of individual joint probabilities, consider the specific outcome where both Alice and Bob obtain $+$. In the spatial scenario, the joint probability is 
\[
\mathbb{P}_{\mathrm{space}}(+,+|1,2) = \Tr(\rho_{AB}\Pi_1^+\otimes \Pi_2^+) = \frac{7}{20}\, .
\]
In contrast, the corresponding temporal joint probability is
\[
\mathbb{P}_{\mathrm{time}}(+,+|1,2) = \Tr\Big(\E\big(\Pi_1^+\rho_A\Pi_1^+\big)\Pi_2^+\Big) = \frac{1}{3}\neq \frac{7}{20}\, .
\]
Therefore, $\mathbb{P}_{\mathrm{space}}(+,+|1,2) \neq \mathbb{P}_{\mathrm{time}}(+,+|1,2)$, allowing one to unambiguously distinguish between the spatial and temporal scenarios from the joint probability data.
\ex

\section{Concluding Remarks}

We have shown that while the correlation data encoded by a dual state render the associated spatial and temporal scenarios as causally indistinguishable, this ambiguity is generally an artifact of classical post-processing. In particular, by analyzing the underlying joint probability distributions which give rise to such correlation data, we demonstrated that the temporal scenario inherently violates the no-signalling condition for any oblique initial state, thus eliminating the quantum causal confusion. Moreover, as oblique states constitute an open dense subset of the Bloch ball, elimination of quantum causal confusion is the norm rather than an exception. 

While our exposition focused on single-qubit systems for clarity, the notion of a dual state, obliqueness, and the elimination of quantum causal confusion immediately generalizes to $n$-qubit systems. An important operational feature of this generalization is that the elimination of causal confusion does not require fine-grained, rank-1 von~Neumann measurements on the global system. Instead, the spatial and temporal scenarios can be distinguished at the level of coarse-grained, binary Pauli measurements yielding outcomes of $\pm 1$. Since the measurement back-action of such coarse-grained measurements is mathematically analogous to the single-qubit case, the temporal violation of the no-signalling condition persists even when Alice and Bob extract only a single bit of global information.

It is interesting to note that for temporal scenarios associated with state-channel pairs of the form $(\mu,\E)$ with $\mu$ the maximally mixed state, the joint temporal probabilities necessarily satisfy no-signalling. Therefore, if $\E$ is not a PPT channel, then the associated pseudo-density operator $\varrho_{AB}=\E\star \mu$ is not a dual state, thus the correlations encoded by $\varrho_{AB}$ may not be realized by a spatial scenario. As such, it follows that there exist joint probability distributions satisfying no-signalling that can only be realized in a temporal scenario, demonstrating that the absence of signalling does not guarantee a common-cause explanation.

\emph{Acknowledgments.---} J.F. is supported by the National Natural Science Foundation of China (W2632013), and by the Hainan Provincial Natural Science Foundation of China (126MS0010). J.F. would also like to thank Arthur J. Parzygnat for fruitful discussion, and for pointing out that the absence of signalling does not guarantee a common-cause explanation. W.Y. is supported by National Natural Science Foundation of China (62475062), and the Humboldt Research Fellowship Programme for Experienced Researchers (CHN-1218456-HFST-E).

\emph{AI usage declaration.---} During the preparation of this manuscript, the authors utilized Google Gemini for language refinement, and as an interactive tool for the exploration of analytical methods and mathematical derivations. The authors rigorously reviewed and edited all generated text and assume full responsibility for the content and technical accuracy of this work.

\bibliography{references}

\end{document}